\documentclass{cernepprep} 
\usepackage{cernunits,cernchemsym,heppennames2}
\usepackage{cite,fancyvrb}

\usepackage{graphicx}
\usepackage{dcolumn}
\usepackage{bm}
\usepackage[utf8]{inputenc}
\usepackage[T1]{fontenc}
\usepackage{mathptmx}
\usepackage{etoolbox}

\begin{document}

\begin{titlepage}
\EPnumber{2022-069}
\EPdate{\today}
\DEFCOL{CDS-Library}

\title{Cyclotron cooling to cryogenic temperature in a Penning-Malmberg trap with a large solid angle acceptance}

\begin{Authlist}
 
C.~Amsler$^{a}$,
H.~Breuker$^{b}$,
S.~Chesnevskaya$^{a}$,
G.~Costantini$^{c,d}$,
R.~Ferragut$^{e,f}$,
M.~Giammarchi$^{f}$,
A.~Gligorova$^{a}$,
G.~Gosta$^{c,d}$,
H.~Higaki$^{g}$,
E.~D.~Hunter$^{a}$\thanks{Electronic mail: eric.david.hunter@cern.ch},
C.~Killian$^{a}$,
V.~Kletzl$^{a}$,
V.~Kraxberger$^{a}$,
N.~Kuroda$^{h}$,
A.~Lanz$^{a,i}$,
M.~Leali$^{c,d}$,
V.~M\"ackel$^{a}$,
G.~Maero$^{f,j}$,
C.~Mal\-bru\-not$^{k}$\thanks{Permanent address: TRIUMF, Vancouver, BC V6T 2A3, Canada},
V.~Mascagna$^{c,d}$,
Y.~Matsuda$^{h}$,
S.~Migliorati$^{c,d}$,
D.~J.~Murtagh$^{a}$,
Y.~Nagata$^{l}$,
A.~Nanda$^{a,i}$,
L.~Nowak$^{k,i}$,
E.~Pasino$^{f,j}$,
M.~Romé$^{f,j}$,
M.~C.~Simon$^{a}$,
M.~Tajima$^{m}$,
V.~Toso$^{e,f}$,
S.~Ulmer$^{b}$,
L.~Venturelli$^{c,d}$,
A.~Weiser$^{a,i}$,
E.~Widmann$^{a}$,
T.~Wolz$^{k}$,
Y.~Yamazaki$^{b}$,
J.~Zmeskal$^{a}$

\end{Authlist}

\Collaboration{(The ASACUSA-Cusp Collaboration)}

$^{a}$Stefan Meyer Institute for Subatomic Physics, Austrian Academy of Sciences, Vienna, Austria, 
$^{b}$Ulmer Fundamental Symmetries Laboratory, RIKEN, Saitama, Japan,
$^{c}$Dipartimento di Ingegneria dell'In\-formazione, Universit\`a degli Studi di Brescia, Brescia, Italy
$^{d}$INFN sez. Pavia, Pavia, Italy,
$^{e}$L-NESS and Department of Physics, Politecnico di Milano, Como, Italy,
$^{f}$INFN sez. Milano, Milan, Italy,
$^{g}$Graduate School of Advanced Science and Engineering, Hiroshima University, Hiroshima, Japan,
$^{h}$Institute of Physics, Graduate School of Arts and Sciences, University of Tokyo, Tokyo, Japan,
$^{i}$University of Vienna, Vienna Doctoral School in Physics, Vienna, Austria,
$^{j}$Dipartimento di Fisica, Universit\`a degli Studi di Milano, Milan, Italy
$^{k}$Experimental Physics Department, CERN, Geneva, Switzerland,
$^{l}$Department of Physics, Tokyo University of Science, Tokyo, Japan,
$^{m}$RIKEN Nishina Center for Accelerator-Based Science, Saitama, Japan,
$^{n}$Nishina Center for Accelerator-Based Science, RIKEN

\begin{abstract}
Magnetized nonneutral plasma composed of electrons or positrons couples to the local microwave environment via cyclotron radiation. The equilibrium plasma temperature depends on the microwave energy density near the cyclotron frequency. Fine copper meshes and cryogenic microwave absorbing material were used to lower the effective temperature of the radiation environment in ASACUSA's Cusp trap, resulting in significantly reduced plasma temperature.
\end{abstract}

\end{titlepage}

\section{Introduction}
\label{sec:introduction}
Antihydrogen, the simplest neutral system composed entirely of antimatter, is produced by combining positron and antiproton plasmas in Penning traps \cite{amor:02, gabr:02, enom:10} at CERN's antiproton decelerator facility \cite{maur:97}. Spectroscopy of the antiatom's microwave \cite{ahma:17a} and optical \cite{ahma:18a} transitions probes CPT-invariance \cite{bluh:99}. Free-fall experiments \cite{kell:08, pere:15, bert:18} will leverage the neutrality of the antiatom to directly test the applicability of the equivalence principle to antimatter. Such experiments may offer insight into the observed asymmetry between matter and antimatter in the universe \cite{dine:03}.

The success of these experiments requires cooling the antimatter plasma to the lowest possible temperature \cite{ahma:17b}. In general, the colder the plasma, the colder will be the antihydrogen produced \cite{jons:18}. Magnetized nonneutral plasma composed of electrons or positrons cools by emitting cyclotron radiation at frequencies close to $\omega_c / 2\pi = 28\,\mathrm{GHz} \times B[\mathrm{T}]$. Cyclotron cooling usually proceeds according to Newton's law of cooling
\begin{equation}
\label{eq:dTdt}
    dT/dt = -\Gamma(T-T_b) + H
\end{equation}
where the cooling rate $\Gamma = 0.26 \,\mathrm{s^{-1}}\times B[\mathrm{T}]^2$ for electrons, the heating rate $H$ is in $\mathrm{K \, s}^{-1}$, and the plasma temperature $T$ tends to the final temperature
\begin{equation}
\label{eq:Tf}
    T_f=T_b+H/\Gamma.
\end{equation} 
The temperature $T_b$ is the blackbody radiation temperature seen by the plasma and will be discussed shortly. The plasma heating rate $H$ has two main contributions: the damping of plasma modes which are excited by electrical noise on the electrodes, and plasma expansion, which converts the potential energy of the concentrated charges into kinetic energy \cite{dani:15}.

After a short time (typically 10 or 20 seconds) the plasma has passively cooled to $T\approx T_f$ and may be further cooled by evaporative cooling \cite{andr:10} and adiabatic expansion \cite{nati:15}. Such "active cooling" causes unavoidable expansion of the plasma and loss of particles. The total reduction in plasma temperature is typically a factor of 10 or less, depending on the initial temperature, the initial density, and how much loss is acceptable. The change in temperature is also temporary. The plasma quickly warms back up to its equilibrium temperature $T_f$. If active cooling is used to continuously suppress the plasma temperature below $T_f$ (as in Ref.~\cite{ahma:17b}), then cyclotron radiation behaves as a heating rate proportional to the difference between $T_f$ and the suppressed temperature. Thus, even where other cooling methods are considered, it is essential to minimize $T_f$, and consequently both $H$ and $T_b$.

The temperature $T_b$ represents the microwave radiation environment which couples to the cyclotron motion of the plasma. $T_b$ is often assumed to be equal to the trap temperature $T_t$. However, that assumption is only valid when the microwaves produced by the plasma are efficiently absorbed by the trap electrodes, for example when the electrodes form a microwave cavity \cite{evet:16}. When this is not the case, the value of $T_b$, and thus, the final temperature of the plasma $T_f$, becomes difficult to predict. The true value of $T_b$ must involve an average of the temperatures of all possible absorbing surfaces, weighted by how resistive the surface is, how strongly the surface couples to a given microwave mode, and how strongly that mode couples to the collective modes of the plasma in a given geometry \cite{kur:20}. In general, $T_b$ is higher than $T_t$. The discrepancy will likely be greater for traps having fewer absorptive surfaces in the cryogenic region and more solid angle open to room-temperature surfaces. 

In this article, it is shown for the first time that correctly addressing these factors can lead to a dramatic reduction in plasma temperature\textemdash in fact the lowest steady-state temperature yet achieved for positron or electron plasma in a trap with a large opening to room-temperature equipment. This work encompasses several iterative design phases. In each phase, the radiation environment was intentionally modified and the base plasma temperature, which is taken as an index for $T_b$, was recorded in many experimental realizations (for example, by varying plasma parameters, magnetic field, trap temperature, or trap pressure). All of the observations support the central hypothesis that $T_b$ can be lowered by confining microwaves to the cryogenic part of the trap.

Section~\ref{sec:geometry} introduces the apparatus, with emphasis on the successive modifications to the microwave environment and the accompanying changes in final plasma temperature. The representative plasma temperatures given in Section~\ref{sec:geometry} are placed in context in Section~\ref{sec:temperature}, where the dependence of plasma temperature on number of electrons $N$ is given for each configuration of the experiment. This section also presents scans over trap temperature $T_t$ and magnetic field $B$ which validate the cooling model presented above and illustrate the connection between $T_f$, $T_b$, and $T_t$. Section~\ref{sec:discussion} details the remaining differences between the configurations which were not mentioned earlier because they do not affect microwave impedance. It is shown that most of these differences cannot contribute to the effect reported here. For the few details that could affect the plasma temperature, the most plausible mechanism is once again a small change in $T_b$, so the argument given above is unchanged. This view is further strengthened via comparisons with other Penning traps where charged particles have been cooled to cryogenic temperature. Section~\ref{sec:conclusion} is the conclusion.

\section{Trap Geometry}
\label{sec:geometry}

\begin{figure}
\centering
    \includegraphics[width=\linewidth]{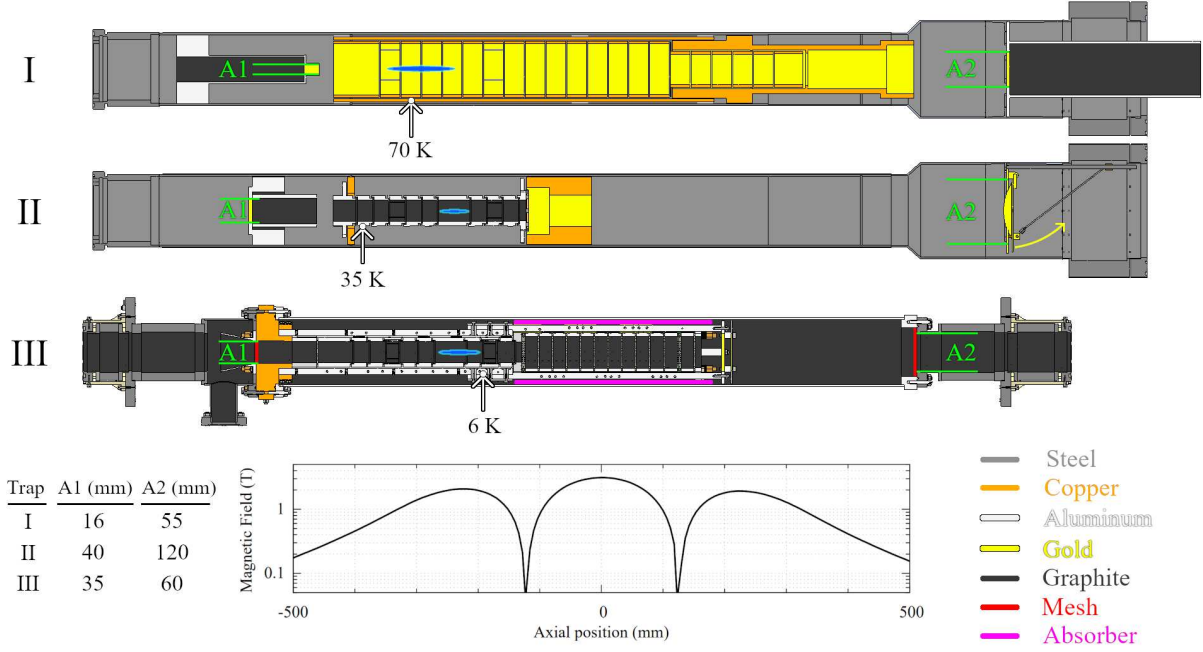}
    \caption{Sketch of the traps and Cusp magnetic field. Left is upstream, right is downstream. Some aluminum structures have been omitted for clarity; these do not intercept the paths of microwaves going out of the trap. Materials having an influence on cryogenics or microwave propagation have been color coded as indicated in the key. The bright blue ellipses float at the approximate position of the plasma.}
    \label{fig:apparatus}
\end{figure}

The plasma studied in this work was confined near the center of the upstream magnetic mirror of ASACUSA's Cusp trap \cite{kuro:17}. As shown in Fig.~\ref{fig:apparatus}, the magnetic field may vary by $10\%$ or more over the length of the plasma. In contrast, an ideal Penning-Malmberg trap would have a perfectly homogeneous magnetic field. In spite of this discrepancy, the plasma behavior relevant to the present work is essentially the same as in a standard trap. The plasma can be compressed with the standard rotating wall technique \cite{sait:08}. The plasma cools at the expected rate following Eq.~\ref{eq:dTdt} (see Section~\ref{sec:temperature}). Expansion rates are similar to those of typical high-field Penning-Malmberg traps (i.e. 1000 seconds or more for the plasma radius to double) \cite{hunt:22}, and the diocotron instability is sometimes present but rarely pronounced. Thus, the phenomena reported below are considered to be of general relevance to plasma cooling in Penning-Malmberg traps.

Figure~\ref{fig:apparatus} also shows how the microwave radiation background was modified in Traps I, II, and III. The primary method was to remove or replace the attenuating or reflecting structures at the openings of the trap (A1 and A2 in Fig.~\ref{fig:apparatus}). The other method was to change the position of a movable radiation shield, which could be opened up to $90^\circ$ as indicated by the yellow arrow on the right side of Trap II in Fig.~\ref{fig:apparatus}. The shield was made of copper and electroplated with gold.

Trap III was designed specifically to prevent the plasma from coupling to regions outside the trap. For Trap III, A1 and A2 were covered with a fine copper mesh (wire diameter $0.03 \,\mathrm{mm}$, pitch $0.25\,\mathrm{mm}$) sprayed with an alcohol-based colloidal graphite anti-static solution. Sprayed samples were inspected under a microscope to ensure that the spraying procedure did not reduce the transparency of the mesh. Theoretically \cite{anke:13}, such a mesh reduces transmitted microwave power at frequencies below $60\,\mathrm{GHz}$ (the cyclotron frequency at the highest magnetic field in the trap) by at least $20\,\mathrm{dB}$. Trap III also contained long sections of resistively coated ceramic rod (``Absorber'' in Fig.~\ref{fig:apparatus}), which had previously been employed for absorbing microwaves in accelerator beamlines by CERN's RF group \cite{casp:93}. These sections were tied to the 5N aluminum bars used for cooling the electrodes and protected from line-of-sight (infrared) radiation from outside; thus, they probably cooled to a temperature close to that of the bars ($6\,\mathrm{K}$). 

Table~\ref{tab:traps} summarizes how the changes in trap geometry affected the amount of non-cryogenic surface seen by the plasma (expressed as relative solid angle) and the lowest achievable final plasma temperature $T_f$ for a plasma containing $N\approx 2\times 10^6$ electrons.

\begin{table}
\caption{Summary of the major differences affecting the radiation environment for Traps I, II, and III. The minimum plasma temperature for $N\approx 2 \times 10^6$ is also given for each configuration. $\Omega/4\pi$ is the fractional solid angle seen by the plasma for surfaces external to the cryogenic trap (defined by A1 and A2 in Fig.~\ref{fig:apparatus}). Trap II is the version shown in Fig.~\ref{fig:apparatus}, while for Trap II.a the diameter A1 was reduced by a factor of two. ``Shield'' refers to the state of the movable radiation shield (for Trap II and Trap II.a). Considering only microwave propagation, $\Omega$ should be close to 0 in Trap III because the apertures are covered by meshes.}
\centering
\renewcommand{\arraystretch}{1.2}
\begin{tabular}{|c|c|c|c|c|}
 \hline
Trap    &   $\Omega/4\pi$ ($10^{-4}$) &   Shield & Trap $T$ (K)	& Plasma $T$ (K)\\  \hline
I   & 11	& -      & 70	& 130 	\\
II   & 20	& open      & 35	& 170 	\\
II    & 11	& closed    & 35	& 150 	\\
II.a    & 12	& open      & 35	& 130 	\\
II.a    & 3	& closed    & 35	& 110 	\\
III   &   11	(or 0) & -	& 6	    & 25 	\\
 \hline
 \end{tabular}
\label{tab:traps}
\end{table}

Plasma temperature measurements will be treated in more detail in Section~\ref{sec:temperature}. However, the most significant trends may already be inferred from the data in Table~\ref{tab:traps}. Increasing the solid angle to surfaces outside the trap (Trap I $\rightarrow$ Trap II) raised the base plasma temperature, and reducing the solid angle (Trap II $\rightarrow$ Trap II.a) reduced the temperature. Likewise, closing the movable shield reduced both solid angle and plasma temperature. Reducing the trap temperature from $70$ to $35\,\mathrm{K}$ (Trap I $\rightarrow$ Trap II) was not correlated with any significant change in plasma temperature, while the reduction from $35$ to $6\,\mathrm{K}$ (Trap II $\rightarrow$ Trap III) was correlated with the most dramatic reduction in plasma temperature. The latter correlation seems to be a coincidence; plasma temperature was still by far the lowest in Trap III even when the trap was warmed to $35\,\mathrm{K}$, as will be described in the next section.

\section{Plasma Temperature}
\label{sec:temperature}

The plasma temperature was measured by slowly releasing the charges onto a microchannel plate (followed by a phosphor screen) and correlating the time-dependent plasma current\textemdash measured by a silicon photomultiplier via the light emitted from the phosphor screen \cite{hunt:20}\textemdash with the time-dependent confinement potential in the trap \cite{eggl:92}. Knowing the fraction of particles which can escape as a function of confinement potential, one can reconstruct the high-energy tail of the Maxwell-Boltzmann distribution for the plasma. The temperature is then estimated by fitting the tail of the distribution to the form
\begin{equation}
\label{eq:max}
    A(t)=A_0 \,\mathrm{exp}[-U(t)/k_B T]
\end{equation}
where $A(t)$ is proportional to the plasma current and $U(t)$ is the energy of the escaping charge with respect to the bottom of the confining well. $U(t)$ decreases as the well is opened on the upstream side. 

\begin{figure}
    \centering 
    \includegraphics[width=0.75\linewidth]{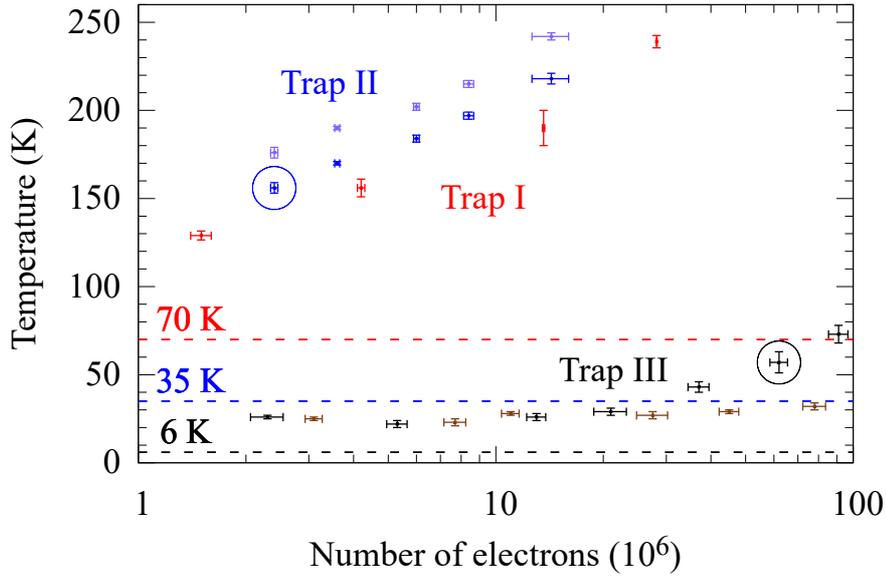}
    \caption{Measured final temperatures for plasma with variable number of electrons $N$. Dashed lines indicate the temperature of the electrodes for each trap version. For Trap I, two data points are off scale at (58, 339) and (115, 454). For Trap II, the pale points correspond to the downstream thermal shield being open. For Trap III, the black points refer to a plasma preparation yielding a relatively higher density than for the brown points. The two encircled points correspond to a plasma density $n\approx 1.0\times 10^8\,\mathrm{cm}^{-3}$.}
    \label{fig:TvsN}
\end{figure}

Figure \ref{fig:TvsN} shows the final plasma temperature $T_f$ obtained for a range of values of $N$, the number of electrons in the plasma. The plasma temperature is reported for four different trap configurations, corresponding to Traps I, II (movable shield open and closed), and III in Table~\ref{tab:traps}. For the data sets shown, the plasma radius and length varied in the range $0.2<r_p<2.5\,\mathrm{mm}$ and $1<L_p<12\,\mathrm{cm}$, respectively. Most of the variation in radius and length is due to the Trap I data set, for which the plasma density was fixed at $n\approx 4 \times 10^8 \,\mathrm{cm}^{-3}$. For the other data sets, the radius was the parameter held constant ($0.7\,\mathrm{mm}$ for Trap II, $1.9\,\mathrm{mm}$ for the black points of Trap III), so that density and length varied in the same sense as $N$ (lower $N$ resulting in lower density and length).

For all values of $N$, the final plasma temperature was lowest for Trap III, in which $2<N<30$ million electrons were cooled to $T_f\approx 25\,\mathrm{K}$. This was the only trap in which the plasma temperature came within $20\,\mathrm{K}$ of wall temperature. The black, red, dark blue, and light blue data sets in Fig.~\ref{fig:TvsN} correspond to progressively higher plasma temperature and more unscreened aperture looking out of the trap ($\Omega$ in Table~\ref{tab:traps}, noting that for Trap III, $\Omega$ should be 0 when the meshes are taken into account). This correlation is monotonic in Fig.~\ref{fig:TvsN}, in spite of the reversal in trap temperature (Trap II was colder than Trap I). In this sense, the data in Fig.~\ref{fig:TvsN} demonstrates the robustness of the conclusions drawn from Table~\ref{tab:traps} over a broad range of plasma parameters.

For Trap II, the brighter points (which correspond to the movable shield being open) are consistently $20\,\mathrm{K}$ higher than the darker blue points. Experiments with the plasma held in different parts of the trap (reported in Ref.~\cite{hunt:22}), where the cooling rate and final temperature were in general different, consistently reproduced this approximately $20\,\mathrm{K}$ offset when the shield was open. In every case where a comparison has been made, the sole effect of opening the shield was to increase $T_b$ by about $20\,\mathrm{K}$. 

For Traps I and II, $T_f$ increases with $N$, whereas for Trap III this behavior is only apparent at high $N$. For Trap I the strong-drive rotating wall technique \cite{dani:07} was employed to fix the density $n\approx 4\times 10^8 \,\mathrm{cm^{-3}}$ for all values of $N$, while for Traps II and III, the cutting method was employed to reduce $N$ without significantly altering the radius of the plasma. $N$ was reduced by stretching the plasma axially and then cutting off a portion and discarding it. This procedure has the side effect of reducing the plasma density, which was relatively low even for the highest $N$ points for Trap III. Thus, the relative flatness of most of the Trap III data set may be the result of a lower overall plasma density. This effect is a coincidence of the plasma parameters chosen and does not affect the main conclusion, as can be seen by comparing the two points enclosed by circles. Both points correspond to a density $n\approx 1.0 \times 10^8\,\mathrm{cm}^{-3}$. For $2.4\times 10^6$ electrons in Trap II, the final temperature was $156\,\mathrm{K}$. The corresponding temperature in Trap III was nearly $100\,\mathrm{K}$ lower ($T_f=57\,\mathrm{K}$), for a plasma containing 25 times more electrons.

\begin{figure}
    \centering 
    \includegraphics[width=0.75\linewidth]{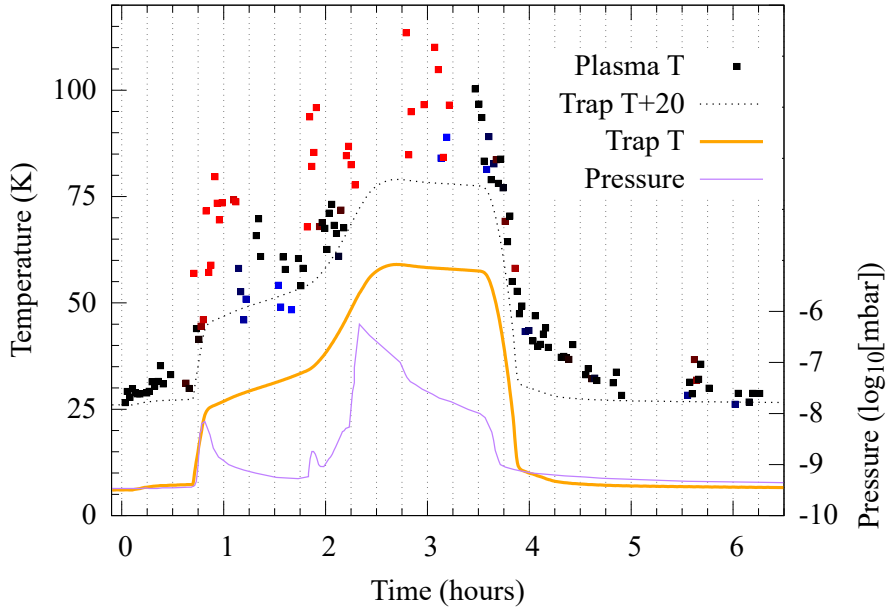}
    \caption{Plasma temperature measurements while Trap III was warmed up and cooled back down. Trap temperature and pressure (outside the trap) were monitored simultaneously. At $t=0$, both coldheads were running. At $t=0.10$, the DS coldhead was stopped. At $t=0.66$, the DS coldhead was started and the US coldhead was stopped. At $t=1.8$ the DS coldhead was stopped. At $t=2.8$ the DS coldhead was started. At $t=3.5$ the US coldhead was started. Point coloring is explained in the text.}
    \label{fig:TfvsTt}
\end{figure}

Figure~\ref{fig:TfvsTt} shows the time evolution of typical plasma temperature as the temperature of Trap III was varied by turning the two coldheads on and off. The upstream (US) coldhead was directly connected to the bars which cool the electrodes, and therefore had a more immediate effect on the measured trap temperature than the downstream (DS) coldhead. As the trap warmed up, gases frozen onto cryogenic surfaces in the trap were liberated and increased the measured pressure. The pressure was measured in a room temperature region external to the trap. During periods of strong off-gassing, the pressure in the trap must have been even higher than the pressure measured outside.

The initial plasma parameters were $T=25\,\mathrm{K}$, $N=27 \times 10^6$, $n \approx 1.7 \times 10^8 \,\mathrm{cm}^{-3}$, $r_p=1.0\,\mathrm{mm}$, and $L_p\approx 5\,\mathrm{cm}$. The plasma parameters changed as the trap warmed and off-gassed because emission from the electron source can change at higher pressure; in particular, the BaO cathode is poisoned by hydrogen. When the gas load was relatively high, $N$ and $r_p$ were found to increase significantly (factor of 2 to 6); this was also correlated with higher temperature. In response, the electron load was tuned to bring the number of electrons and plasma radius close to their original value. To give some indication of these deviations, the points on Fig.~\ref{fig:TfvsTt} are color coded by the space charge. This is the value of the vacuum well depth at the moment that the first particles begin to escape, and it correlates positively with the number of electrons in the plasma. Black points correspond to plasma with a space charge between $6$ and $8$ volts. Blue points mean the plasma was smaller, i.e. space charge $< \,6\,\mathrm{V}$; red points mean the plasma was larger. The black points more or less follow a minimum-temperature curve which is consistently $20\,\mathrm{K}$ higher than the trap temperature $T_t$. An interesting exception to this generalization occurred at $t=4$ hours, where the black points fall more slowly than the measured trap temperature. This could be a sign that the plasma was coupling to something (for example, the weakly coupled ceramic absorbing material) which did not cool down as quickly as the 5N aluminum bar housing the temperature sensor.

\begin{figure}
    \centering 
    \includegraphics[width=0.75\linewidth]{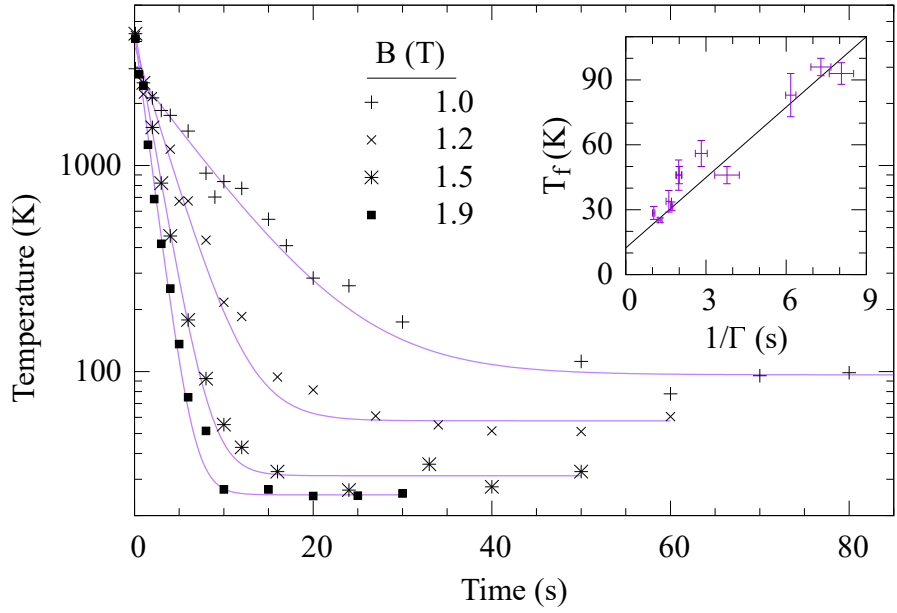}
    \caption{Cooling curves for different values of the magnetic field. The plasma parameters are the same as for Fig.~\ref{fig:TfvsTt}. The plasma temperature falls initially as $T\sim \mathrm{exp}[-\Gamma t]$ and asymptotes to a final temperature $T_f$. The fit curves are used to determine $\Gamma$ and $T_f$. Eleven such fit-parameter pairs are plotted in the inset. The trendline has slope $H$ and $y$-intercept $T_b$, following Eq.~\ref{eq:Tf}.}
    \label{fig:TvsB}
\end{figure}

The plasma may be heated by applying broadband radiofrequency noise to an electrode \cite{kurz:96}. Provided that the noise is applied for a time which is long compared to the collisional relaxation time, the plasma temperature can be raised reproducibly to any value up to a few $\mathrm{eV}$. When the noise is switched off, the plasma cools back down toward $T_f$ according to Eq.~\ref{eq:dTdt}. One can measure $\Gamma$ by fitting the slope (on a semi-log plot) of the cooling curve, $T$ vs. $t$, where $t$ is the time between turning off the noise and measuring the plasma temperature. Such measurements have been performed extensively in Traps II and III. Typical results for Trap III are shown in Fig.~\ref{fig:TvsB}.

The cooling rate $\Gamma$ depends on the magnetic field $B$, and the latter can be reduced by ramping down the current in the Cusp Magnet. By repeating the cooling curve measurement for different values of $B$, 11 pairs of fit parameters \{$\Gamma$, $T_f$\} were obtained. These are shown in the inset to Fig.~\ref{fig:TvsB}. Using Eq.~\ref{eq:Tf}, the data is fit by a line with slope $H=11\pm 1\,\mathrm{K/s}$ and $y$-intercept $T_b=12\pm 2\,\mathrm{K}$. However, the accuracy of this result depends on the approximation that $H$ is constant when $B$ is varied. In fact, lower $B$ implies faster plasma expansion. The additional expansion heating would make $H$ a function of $1/\Gamma$, so that the data would curve upward instead of following a line. In this situation, a linear fit tends to overestimate $H$ (for small $1/\Gamma$) and underestimate $T_f$. 

The tendency for faster plasma expansion at low $B$ was intensified by the fact that the Cusp field's symmetry axis also depends on $B$. This is likely a combination of two effects: (1) nonlinearity due to magnetic saturation in the hot-rolled steel shield enclosing the Cusp Magnet and (2) a different distribution of supercurrents in the anti-Helmholtz coils chosen to reach lower $B$ at the plasma. The angle between the symmetry axes of $\mathbf{E}$ and $\mathbf{B}$ was tuned to be less than $1\,\mathrm{mrad}$ for $B=1.9\,\mathrm{T}$. The alignment was not retuned when $B$ was changed, resulting in a misalignment of approximately $3\,\mathrm{mrad}$ for the lowest $B$ value studied. The plasma expansion time was $8500\,\mathrm{s}$ for the highest value of $B$ and $900\,\mathrm{s}$ for the lowest, corresponding to expansion heating rates $1\,\mathrm{K/s}$ and $12\,\mathrm{K/s}$, respectively. To further complicate matters, the slowest cooling was observed at an intermediate field value where cooling was suppressed well below the free-space value, presumably by cavity-resonant effects. This explains why the constant best-fit value $H=11\,\mathrm{K/s}$ does not fit the data perfectly.

In view of these limitations, the trendline in Fig.~\ref{fig:TvsB} does not provide a reliable estimate of either the heating rate $H$ or the base temperature $T_b$. However, the fit parameters are useful for providing bounds on these quantities. In particular, supposing that $H$ does not decrease when $B$ is decreased, then $H$ cannot be much more than the fitted value of $11\,\mathrm{K/s}$; this agrees with the heating rate estimated in the preceding paragraph from the expansion rate at low $B$. This upper bound on $H$, combined with the high-field points at $1/\Gamma \approx 1\,\mathrm{s}$ and $T_f\approx 25\,\mathrm{K}$, allows a lower bound to be placed on $T_b$ at the fitted value $T_b=12\pm 2\,\mathrm{K}$. Meanwhile, an upper bound for $T_b$ is the minimum observed $T_f\approx 25\,\mathrm{K}$. These measurements therefore establish that $12 \lesssim T_b < 25\,\mathrm{K}$ in Trap III.

\section{Discussion}
\label{sec:discussion}

The time dependence of the plasma temperature in Trap III is well modeled by Newton's law of cooling, Eq.~\ref{eq:dTdt}, with an effective background temperature $T_b\approx 20\,\mathrm{K}$. Of the three trap designs studied, only Trap III, which was engineered to contain the cyclotron radiation of the plasma, could produce plasma with $T_f$ anywhere near the trap temperature $T_t$. Cooling curves taken in Traps I and II asymptote to much higher $T_f$, implying higher effective background temperature $T_b$ (for equivalent $H$). For example, in Ref.~\cite{hunt:22} a measurement similar to Fig.~\ref{fig:TvsB} was performed in Trap II and yielded $T_b\sim 100\,\mathrm{K}$.

The best explanation for these observations is that the copper meshes in Trap III reflected radiation at the cyclotron frequency, confining cyclotron radiation to the coldest part of the trap and fixing $T_b$ at the temperature of the cryogenic absorbing material. However, the meshes were not the only feature distinguishing Trap III from the others. The following subsections review the remaining differences among the traps and compare the results of the present work to those found in similar traps elsewhere.

\subsection{Other differences between the traps}
\label{sec:differences}

Noise on the electrodes can heat the plasma. Noise was minimized for all configurations by using low-speed amplifiers, low-pass filters enclosed in a thick aluminum box mounted directly to the vacuum feedthrough, and cryogenic low-pass filters close to the electrodes. For Trap I, the cryogenic filters were bypassed by diodes in order to allow fast high-amplitude signals to pass. However, because this design also admits noise any time the electrode bias changes in a time comparable to the filter time constant, the diode bypass was removed in Traps II and III for all but the outermost electrode (which is used for catching and therefore must be pulsed). For Trap I, signals were carried through vacuum a distance of approximately $1.0\,\mathrm{m}$ on homemade stainless steel coaxial cables, with the outer shield grounded at the trap side. For Trap II, the same distance was spanned by Lakeshore Quad-Twist\textsuperscript{TM} phosphor bronze twisted pair wire, with every other wire grounded at both sides. For Trap III, more Quad-Twist\textsuperscript{TM} wire was used, with the distance traversed in vacuum reduced by a factor of two.

The inner diameter of the electrodes in Trap I was $80\,\mathrm{mm}$, while the inner diameter of the electrodes in Traps II and III was $34\,\mathrm{mm}$. The electrodes of Trap I were gold-coated, while the electrodes of Traps II and III were coated first in gold, then in colloidal graphite (Electron Microscopy Systems \#12660). The inner surfaces of the cold bore housing Trap III were also darkened with the alcohol-graphite solution, whereas some of the surfaces for Traps I and II were highly reflective.

Trap III was installed as part of a new cold bore with two new coldheads. The electrode stacks of Trap I and II were placed in an older cold bore with similar dimensions, but with the downstream coldhead inactive, so that the temperature of the vacuum chamber was higher in general and in particular increased further downstream, approaching $200\,\mathrm{K}$ in the vicinity of the downstream aperture. This may account for the smallness of the effect of closing the movable shield in Trap II. It would also imply that cryopumping was much less effective in Traps I and II. However, the latter may have been offset by the fact that Trap III was only pumped on the upstream side, whereas Traps I and II were pumped from upstream and downstream by non-evaporable getter pumps, with the typical pressure at the downstream side being less than $3\times 10^{-10}\,\mathrm{mbar}$.

One may now consider whether any of these differences could lead to an alternative explanation for the lower plasma temperature observed in Trap III. The use of twisted pair wiring, filterboards without diodes, smaller electrodes, and colloidal graphite coating were common factors between Trap II and Trap III, and can therefore be ruled out. Any heating mechanism based on vacuum quality or cryopumping is on very weak ground, considering the relatively low temperatures observed in Fig.~\ref{fig:TfvsTt} in the presence of a warmer trap and a relatively poor vacuum. It is also difficult to construct a consistent argument for more residual gas-related heating in Trap II than Trap III, given that the plasma expansion rate was negligibly low in Trap II \cite{hunt:22}. The only differences which may have affected plasma temperature were the inactive coldhead and lack of graphite coating on some cryogenic surfaces in Traps I and II. However, if these differences did contribute to plasma heating, the only plausible heating mechanism seems to be, once again, a change on the effective background temperature $T_b$. 

\subsection{Cooling in other traps}
\label{sec:other}
Coupling to room-temperature radiation is a general problem for the cooling of charged particles in cryogenic open-ended traps. Temperatures as low as those reported here are seldom observed without applying active techniques such as laser or evaporative cooling. The few exceptions involve traps where room temperature microwave radiation was reduced, excluded from the cryogenic region, or otherwise rendered insignificant:
\begin{enumerate}
    \item The BASE collaboration achieved sub-kelvin cyclotron energy for a single antiproton in a Penning trap with a $2\,\mathrm{T}$ magnetic field \cite{smor:17}. The trap was entirely enclosed by surfaces cooled to $6\,\mathrm{K}$ and sealed with indium. Similarly, a small Penning trap ($1\,\mathrm{cm^3}$ surrounded by electrodes), nominally at $1.6\,\mathrm{K}$ in a $5\,\mathrm{T}$ field, was used to cool a single electron to $5\,\mathrm{K}$ via cyclotron radiation. The cyclotron motion of the electron was further cooled to $0.85\,\mathrm{K}$ by suppressing the cyclotron coupling to the electrodes (detuning from a cavity resonance) and using feedback from a cryogenic amplifier \cite{durs:03}.
    \item The thesis of Beck \cite{beck:90} reported plasma containing $N\sim 10^7$ electrons cooled as low as $T\lesssim 30\,\mathrm{K}$. As in the preceding example, the trap was pinched off and indium sealed. The trap was immersed in helium and operated at $6\,\mathrm{T}$.
    \item The ALPHA collaboration reported the cooling of plasma containing $2.6\times 10^6$ positrons to $T\approx 50\,\mathrm{K}$ via cyclotron radiation in a $3\,\mathrm{T}$ magnetic field \cite{ahma:17b} (more recent measurements \cite{bake:21} suggested a lower temperature $T\sim 20\,\mathrm{K}$, however this temperature was measured after an adiabatic cooling step and therefore does not describe the steady state of the plasma). The helium-cooled trap had open endcaps in order to permit axial transport of electrons, positrons, antiprotons, and ions. The apertures enclosing the trapping region had a measured temperature of $6\mbox{--}7\,\mathrm{K}$ and had inner diameter $d\approx 10\,\mathrm{mm}$. In addition to these apertures, access for an off-axis laser beam was provided through a long $11\,\mathrm{mm}$ diameter port containing a resistively coated ceramic microwave absorber \cite{evet:15}.
    \item The thesis of Hunter \cite{hunt:19} reported electron plasma cooled to within a few degrees of trap temperature ($9\,\mathrm{K}$) for the entire range $10^4 < N < 10^7$. The trap was open at both ends with $10\,\mathrm{mm}$ copper apertures. The plasma was cooled using cyclotron-cavity resonance in a $1\,\mathrm{T}$ field. On-resonance, the coupling to the cavity mode was typically twenty times higher than the Larmor rate, which limited the influence of modes external to the trap by a similar factor.
\end{enumerate}


\noindent The antihydrogen groups at CERN (AEgIS, ALPHA, ASACUSA, GBAR) cannot perform their experiments in a hermetically sealed cryogenic trap. The results obtained so far by the ALPHA collaboration appear to be a compromise between low temperature and finite acceptance. Groups like AEgIS and ASACUSA cannot make the same compromise as ALPHA, for they require a large opening from the cryogenic region in order for antihydrogen atoms to escape. These groups have struggled for many years to achieve sufficiently cold plasma, and may ultimately conclude that it is impossible to do so without blocking room temperature microwave radiation with methods similar to those presented here. The ALPHA and GBAR groups might also benefit from the improved access and shielding provided by a larger aperture covered in a mesh which blocks microwaves. In contrast to a simple copper mesh, the absorber used for ALPHA's laser port has a limited range of application due to its large aspect ratio and small diameter. Further, the device did not attenuate at sufficiently high frequency for cooling in magnetic fields greater than $1\,\mathrm{T}$.

\section{Conclusion}
\label{sec:conclusion}

The final temperature of the electron plasma held in ASACUSA's Cusp trap was anomalously high (Trap I) and even increased after a significant reduction in trap temperature (Trap II). The addition of fine copper meshes (Trap III) provided an effective barrier against microwaves entering the trap at the cyclotron frequency and coincided with a dramatic reduction in plasma temperature.

The effect was found to be independent of many variations in plasma parameters. A movable thermal shield permitted a partial study of the apparent heating in a system without meshes. The plasma was consistently $20\,\mathrm{K}$ hotter with the shield open, for $3<N<14$ million electrons. Earlier work had found that the temperature difference was the same for four different plasma positions (with different cooling and expansion rates). For the system with meshes, a consistent reduction of about $100\,\mathrm{K}$ was witnessed over a range of plasma density (factor of 4), radius (factor of 2), and number of electrons (factor of 50).

The results of the present study offer strong support for the hypothesis that the meshes allow the plasma to cool to lower temperature by blocking radiation that could couple the plasma to warmer regions of the apparatus. This finding enables the formation of magnetized cryogenic electron plasma in a trap with large solid angle acceptance to room-temperature apparatus. There is no reason to expect that a positron plasma would cool differently than the electron plasma used in this study. Cold positrons are a prerequisite for ASACUSA's planned measurement of the hyperfine splitting in ground-state antihydrogen \cite{malb:18}, which requires a large acceptance for a diffuse antihydrogen beam to exit the trap. Theoretically \cite{radi:14}, a positron plasma with properties similar to the electron plasma described here could be used to generate two orders of magnitude more ground-state antihydrogen than has previously been achieved in ASACUSA's Cusp trap.

\section*{Acknowledgments}
The meshes and microwave absorbers used in Trap III were proposed by Fritz Caspers of CERN's RF group. Engineering design and procurement for Trap III were undertaken chiefly by Doris Pristauz-Telsnigg of the Stefan Meyer Institute. This work was supported by Austrian Science Fund (FWF) P 32468, W1252-N27, and P 34438; the JSPS KAKENHI Fostering Joint International Research B 19KK0075; the Grant-in-Aid for Scientific Research B 20H01930;  Special Research Projects for Basic Science of RIKEN; Università di Brescia and Istituto Nazionale di Fisica Nucleare; and the European Union’s Horizon 2020 research and innovation programme under the Marie Skłodowska-Curie grant agreement No 721559.

\section*{Author Declarations}
\subsection*{Author Contributions}
H.H., E.D.H., A.L., V.M\"a., D.J.M., N.K., and M.T. developed and operated the traps used for plasma cooling. E.D.H., V.Kl., A.L., V.M\"a, D.J.M., M.S., and J.Z. developed the new cold bore for Trap III. Trap upgrades were carried out by C.A., H.B., G.C., R.F., M.G., G.G., A.G., E.D.H., A.L., M.L., V.M\"a, G.M., S.M., D.J.M., L.N., M.R., V.T., A.W., E.W., T.W., and J.Z. All authors critically reviewed and approved the final version of the manuscript.
\subsection*{Conflict of Interest}
The authors have no conflicts to disclose.

\section*{Data Availability}
The data that support the findings of this study are available from the corresponding author upon reasonable request.

\end{document}